\documentclass[preprintnumbers,amsmath,amssymb,axodraw,nofootinbib,showkeys,showpacs]{revtex4}

\usepackage{amsmath}
\usepackage{graphicx}
\usepackage{epsf}
\usepackage{psfrag}
\usepackage{epsfig}
\usepackage{graphics}
\usepackage[dvips]{color}
\newcommand{\Dslash}{\hbox{$\partial\!\!\!{\slash}$}}
\newcommand{\Aslash}{\hbox{$A\!\!\!{\slash}$}}

\newcommand{\pslash}{\hbox{$p\!\!\!{\slash}$}}
\newcommand{\bslash}{\hbox{$b\!\!\!{\slash}$}}
\newcommand{\kslash}{\hbox{$k\!\!\!{\slash}$}}
\newcommand{\kappaslash}{\hbox{$\kappa\!\!\!{\slash}$}}
\newcommand{\be}{\begin{equation}}
\newcommand{\ee}{\end{equation}}
\newcommand{\bq}{\begin{eqnarray}}
\newcommand{\eq}{\end{eqnarray}}

\begin{document}

\title{Consistency of an alternative CPT-odd and Lorentz-violating extension of QED}

\date{\today}
\author{J. C. C. Felipe$^{(a,d)}$} \email[] {jean.cfelipe@ufvjm.edu.br}
\author{H. G. Fargnoli$^{(b)}$} \email[]{helvecio.fargnoli@dex.ufla.br}
\author{A. P. Baeta Scarpelli$^{(c)}$} \email[] {scarpelli@cefetmg.br}
\author{L. C. T. Brito$^{(d)}$} \email[]{lcbrito@dfi.ufla.br}

\affiliation{(a) Instituto de Engenharia, Ci\^encia e Tecnologia, Universidade Federal dos Vales do Jequitinhonha e Mucuri  \\
Avenida Um, 4050 - 39.447-790, Cidade Universitária, Jana\'uba, Minas Gerais, Brazil}

\affiliation{(b) Universidade Federal de Lavras - Departamento de Ci\^encias Exatas \\
Caixa Postal 3037, 37.200-000, Lavras, Minas Gerais, Brazil}

\affiliation{(c)Centro Federal de Educa\c{c}\~ao Tecnol\'ogica - MG \\
Avenida Amazonas, 7675 - 30510-000 - Nova Gameleira - Belo Horizonte
-MG - Brazil}

\affiliation{(d) Universidade Federal de Lavras - Departamento de F\'{\i}sica \\
Caixa Postal 3037, 37.200-000, Lavras, Minas Gerais, Brazil}

\begin{abstract}
\noindent
We investigate an alternative CPT-odd Lorentz-breaking QED which includes the Carroll-Field-Jackiw (CFJ) term of the Standard Model Extension (SME), writing the gauge sector in the action in a Palatini-like form, in which the vectorial field and the field-strength tensor are treated as independent entities. Interestingly, this naturally
induces a Lorentz-violating mass term in the classical action. We study physical consistency aspects of the model both at classical and quantum levels.

\end{abstract}



\maketitle

\section{Introduction}
There is no doubt that Lorentz and gauge symmetries have fundamental importance for understanding the interactions between elementary particles. However, although experiments put very strong constraints on the violation of these symmetries, they are not protected by underlying principles. Thus, they might be approximate symmetries from a more fundamental theory which has the Standard Model as an effective description at low energy.

It is commonly believed that a direct consequence of an exact gauge symmetry in QED is that photons must be massless. However, physical effects beyond Maxwell electromagnetism were extensively discussed in the context of de Broglie-Proca model \cite{Robles:2012zz}-\cite{spavieri}. Regarding the effective character of this symmetry, it may be possible that photons have a very small but nonvanishing rest mass $m_\gamma$. Although methods using different astrophysical sources and laboratory experiments put strong limits on the mass of the photon \cite{Wei:2018pyh}-\cite{mass-limit-2}, the possibility of a non-vanishing $m_\gamma$ has been recently discussed by many authors in different contexts, such as a model with Lorentz and supersymmetry breaking \cite{Bonetti:2016vrq,bonettiA}, the photon-photon scattering \cite{ElMenoufi:2012yh}, the regularization of QCD+QED on the lattice \cite{Endres:2015vpi}, the current confinement in $2+1$ dimensions \cite{Vyas:2017vea}, the black hole formation \cite{Emelyanov:2016rwm} and as a possible imprint of strings \cite{Kostelecky:1990pe}, extra-dimensions \cite{Alencar:2015rtc} and inflation \cite{Prokopec:2002jn}.

The possible violation of the symmetries of special relativity at high energy has been considered  by many authors since the original works by Kosteleck\'y and Samuel \cite{Kostelecky1,Kostelecky2}. In fact, very different scenarios for very high energy physics agree that the usual symmetries of special relativity are expected to be broken in this limit as a natural consequence of quantum gravity effects \cite{Kostelecky1,Kostelecky2,String}. Here we are concerned with the possibility of Lorentz-violating modifications in the framework of the Standard Model Extension (SME) \cite{Colladay:1998fq}. In the SME, the Lorentz-violating terms are provided by adding to the minimal Standard Model all possible Lorentz-violating terms that could arise from spontaneous symmetry-breaking at very high energy \cite{Colladay:1996iz}. These terms incorporate constant tensors which emerge from the spontaneous Lorentz symmetry-breaking. The magnitude of these background tensors must be fixed by experiments (see \cite{Kostelecky:2008ts} for experimental results) and, of course, are expected to be very tiny. The model is interpreted as an effective description of Lorentz violation at low energy. The SME preserves $SU(3) \times SU(2)\times U(1)$ gauge symmetry and renormalizability.

The modification of classical electrodynamics by the inclusion of a CPT- and Lorentz-violating term in the photon sector has been discussed originally by  Carroll, Field and Jackiw \cite{Carroll:1989vb}. Initially, the authors suggested that the polarization of radio signs coming from distant astrophysical objects could be an effect of cosmic anisotropies associated with the violation of Lorentz symmetry \cite{Carroll:1997tc}. With the advent of the SME \cite{Colladay:1998fq}, other possible physical systems which may furnish signs of Lorentz violation were proposed \cite{Coleman:1998ti}-\cite{Cas-Man}. Turning back to the particular case of the Carroll-Field-Jackiw (CFJ) term, an important question was first discussed in reference \cite{Jackiw:1999yp}, in which Kosteleck\'y and Jackiw  addressed the question of its quantum induction when an axial Lorentz- and CPT-violating term is added to the fermionic sector. In this paper, we focus on an alternative CPT-odd extension of QED which includes a CFJ term, writing the gauge contributions in the action in the Palatini-like form, where $A^{\mu}$ and $F^{\mu\nu}$ are treated as independent fields \cite{Arnowitt:1962hi}. Interestingly, this naturally induces a Lorentz-violating mass term in the classical action. We, then, investigate the physical consistency aspects of the model both at classical and quantum levels.

This Letter is organized as follows. In section \ref{model}, we present our model and carry out some general discussions.
The question of the spacetime nature of the background vector responsible for the Lorentz symmetry breaking in the CFJ term is addressed in section \ref{sec-propagator}, with the analysis of the propagating modes of the gauge field. In section \ref{sec-quantum} we consider the quantum model. In particular, we present the one-loop calculation of the radiative corrections to the two-point functions of the gauge and fermion fields up to second-order in the Lorentz-violating parameters. Finally, we conclude in section \ref{sec-conclusion}.

\section{General discussions on the model}
\label{model}

The usual classical action for the CPT-odd Lorentz-breaking Electrodynamics from the minimal Standard Model Extension has the form below:
\begin{equation}
S = \int d^4x\left\{\mathcal{L}_{QED} + \frac{1}{2}\kappa^\sigma \epsilon_{\sigma\lambda\mu\nu} A^{\lambda}F^{\mu\nu}
 - b_{\mu}\bar{\psi}\gamma^{\mu}\gamma_{5}\psi\right\},
\label{CPTodd}
\end{equation}
in which the first term is the standard QED Lagrangian density, the second term is the  Carroll-Field-Jackiw modification of the photon sector and, the last one, a Lorentz- and CPT-breaking term in the fermionic sector. The Lorentz-breaking backgrounds $\kappa^{\mu}$ and $b^{\mu}$ are constant axial vectors and  are defined in quantum theory by convenient renormalization conditions. Actually, there are  tight  bounds on the Lorentz-violating parameters of the SME obtained from experiments both at low and high energy \cite{Kostelecky:2008ts}. This action is gauge-invariant, with $F^{\mu\nu}$ being the usual Maxwell field-strength tensor.

In this paper, however, we study an alternative form of the CPT-odd electrodynamics. First, we write the  gauge contributions in the action in a Palatini-like form, taking $A^{\mu}$ and $F^{\mu\nu}$ as independent fields,
\begin{equation}
S_{G}= \int d^4x\left\{ -\frac{1}{2} F_{\mu\nu}\left( \partial^{\mu}A^ {\nu} - \partial^{\nu}A^{\mu}\right) + \frac{1}{4} F_{\mu\nu}F^{\mu\nu} + \frac{1}{2}\kappa^\sigma \epsilon_{\sigma\lambda\mu\nu} A^{\lambda}F^{\mu\nu}\right\}.
\label{Palatini}
\end{equation}
We then use the Euler-Lagrange equations for $F_{\mu\nu}$, obtained from the Lagrangian density in (\ref{Palatini}), to write
\begin{equation}
F_{\mu\nu}=\partial_{\mu}A_{\nu}-\partial_{\nu}A_{\mu}-\kappa^{\sigma}\epsilon_{\mu\nu\sigma\rho}A^{\rho}.
\label{tens-str}
\end{equation}
The field-strength tensor above has an unusual form, taking into account that we are dealing with an Abelian model. In other words, if one does not assume any explicit relationship between $F_{\mu\nu}$ and $A_\mu$, the variation of the action (\ref{Palatini}) with respect to $F_{\mu\nu}$  gives rise to a non-conventional field-strength tensor. In fact, it is clear that $F_{\mu\nu}$ in equation (\ref{tens-str}) is not invariant under the transformation  $A^{\mu}\rightarrow A^{\mu}+\partial^{\mu}\Lambda(x)$, due to the presence of the extra term. Furthermore, using (\ref{tens-str}) in (\ref{Palatini}), we obtain the Lagrangian density for the modified photon sector
\begin{equation}
\mathcal{L}_{G}=-\frac{1}{4}\left(\partial^{\mu}A^{\nu}-\partial^{\nu}A^{\mu}\right)\left(\partial_{\mu}A_{\nu}-\partial_{\nu}A_{\mu}\right)
+\kappa^{\mu}\epsilon_{\mu\nu\rho\sigma}A^{\nu}\partial^{\rho}A^{\sigma}+\frac{1}{2}A^{\mu}M_{\mu \nu}A^{\nu},
\label{mass-lagrang}
\end{equation}
with $M_{\mu \nu}=\kappa^{2}\eta_{\mu\nu}-\kappa_{\mu}\kappa_{\nu}$.

It is important to observe that this gauge sector of (\ref{mass-lagrang}) which will be studied in this article is different from the one of eq. (\ref{CPTodd}). While the action (\ref{CPTodd}) is gauge invariant (although the Lagrangian density is not), the action built with the Lagrangian density (\ref{mass-lagrang}), with the addition of the usual interaction and fermion terms, is not gauge invariant. Here, we have an explicit gauge-breaking term, with a tensor $M_{\mu \nu}$ coupled to the vector field, which we call with some freedom a mass term. This kind of term in works on field theory in curved spacetime is normally called nonminimal, since it accommodates nonminimal couplings to the curvature in these models \cite{Curv-space1}, \cite{Curv-space2}, \cite{Curv-space3}, \cite{Curv-space4}.

A note on the Palatini formulation is in order. Usually, in this formulation each term is split into two, one depending on only one of the fields and another depending on both. This is the case for the Maxwell term, for which we have a term depending only on $F^{\mu \nu}$ and another with $A^\mu$ and $F^{\mu \nu}$. However, this procedure does not work for the CFJ term, since it is not possible to write such a term depending only on $F^{\mu \nu}$. Another possibility would be to use a CFJ with only $A^\mu$, but, in this case, the field equations are not changed and it would be innocuous for the model. However, it is interesting to observe that it is possible to build an arbitrary $\kappa^2$ order term by using, in the place of $F_{\mu \nu}$, the combination $\alpha F_{\mu \nu} +(1-\alpha)(\partial_\mu A_\nu - \partial_\nu A_\mu)$ in the CPT-odd term of eq. (\ref{Palatini}), with $\alpha$ an arbitrary dimensionless constant.

We observe that this Palatini-like formulation, with the participation of the CFJ term, caused the emergence of an unusual mass contribution in the photon Lagrangian density. While the usual de Broglie-Proca mass term, although breaking gauge invariance, respects Lorentz symmetry, the present mass contribution breaks these two symmetries. Besides, it is relevant to enforce the role of the CFJ term in generating this mass for the photon, since if we have $\kappa^\mu=(0,0,0,0)$ the gauge symmetry is restored. Therefore, we have an interesting case in which Lorentz-symmetry violation concur with the breaking of gauge invariance. The breakdown of gauge invariance can cause undesirable problems in the model, as violation of unitarity, with deleterious effects in the renormalizability of the theory. For the usual de Broglie-Proca model, a solution was found by Stueckelberg \cite{stueckelberg,stueckelberg1}, in which a mixed term that includes a scalar field is added to the Lagrangian density, such that gauge symmetry is restored. It happens that the de Broglie-Proca model can be obtained by gauge fixing the Stueckelberg Lagrangian. Although it is out of the scope of this paper, we believe this procedure could also be adopted in the present model. In \cite{lv-mass4}, the authors developed a procedure to extend the Stueckelberg formalism to models with Lorentz violation, including mass terms, by introducing a scalar field $\phi$ such that
\be
\delta{\cal L}_m=\frac 12 \left(\partial_\mu \phi- m A_\mu\right)\hat \eta^{\mu \nu}\left(\partial_\nu \phi- m A_\nu\right),
\ee
where $\hat \eta^{\mu \nu}=\eta^{\mu \nu}+G^{\mu \nu}$, in which $G^{\mu \nu}$  represents the Lorentz-breaking contribution. The present model can be included in this class by using $G^{\mu \nu}$ proportional to $\kappa^\mu \kappa^\nu/\kappa^2$. Another possibility is applying the concept of dual theories, by using, for example, the gauge embedding procedure (see, for example \cite{lv-mass6}, for a model with a similar gauge-symmetry violation). Since one can consider the non-invariant model as the gauge fixed version of a gauge theory, it would be useful in a future study to carry out such an analysis. Hidden symmetries may be revealed by the construction of a gauge invariant theory from a non-invariant one.

This dependence of the photon mass on the Lorentz-breaking parameter is interesting and deserves some discussion. The stringent limits on the magnitude of $\kappa^\mu$, considering only the CFJ term, is being constantly estimated and discussed with the use of a great variety of techniques and experiments. The most restrictive results are due to astrophysical investigations, such that $\left|\vec\kappa\right|\leq 10^{-42} GeV$ \cite{Kostelecky:2008ts}, \cite{Carroll:1989vb}, \cite{Kost-Mewes}, which is near the Heisenberg limit. In \cite{gomesE}, local phenomena and experiments, like the effects of the CFJ term on the energy spectrum of hydrogen and applications on resonant cavities, were used to establish bounds on the magnitude of the background vector. For the absolute value of the spatial component, it was obtained an upper limit of $8 \times 10^{-23}\,GeV$. Concerning the photon mass, recently Particle Data Group \cite{tanabashiC} presented upper limits obtained with estimations from solar wind magnetic field in \cite{ryutovD1} and \cite{ryutovD}, of $5,6 \times 10^{-28} \, GeV$. However, the procedure adopted for achieving this result was considered optimistic in \cite{retinoB} and \cite{bonettiA}. On the other hand, more conservative limits on the photon mass were obtained from observations made by the Pioneer 10 probe \cite{mass-limit-1}, such that $m_\gamma \leq 6 \times 10^{-25}\, GeV$ \footnote{More stringent limits could be considered, based on the properties of magnetic fields on galactic scales \cite{mass-limit-2}. However, such limits were obtained using the assumption that Lorentz symmetry is preserved.}.

Another question that arises is related with charge conservation. This can be dealt with in the same way the simple de Broglie-Proca-model is treated, imposing a gauge condition. In a more general Lorentz-violating context, the mass term is $(1/2)A^\mu M_{\mu \nu} A^\nu$. If one writes the field equation for $A^\mu$, considering the coupling to an electromagnetic current $J^\mu$, the gauge condition,
\be
M_{\mu \nu}\partial^\nu A^\mu=0,
\ee
assures current conservation. All the procedures adopted in the traditional de Broglie-Proca-model in order to show its physical consistency can be generalized here for the model with a Lorentz-breaking mass term.

It is important to note that the usual CFJ model already accommodates an effective mass for the photon which is proportional to the absolute value of the Lorentz-breaking vector \cite{Bonetti:2016vrq}, \cite{bonettiA}. In \cite{Bonetti:2016vrq}, four general classes of supersymmetry and Lorentz symmetry breaking were studied in the context of the SME. These classes include both CPT-odd and CPT-even QED, considering or not the influence of the photino in the photon sector. It was found that dispersion relations exhibit non-Maxwellian behavior. Particularly in the case of CPT-odd QED, the CFJ term was shown to provide the photon a mass which is proportional to the magnitude of the spatial part of the background vector and which preserves gauge invariance. Further, in \cite{bonettiA}, this investigation was expanded and features like complex frequencies, superluminal group velocities and possible nonconservation of the energy-momentum tensor were discussed in detail. In the present study, we restrict ourselves to the investigation of some physical consistency aspects of a CPT-odd model which incorporates a Lorentz-breaking mass term for the photon. Here, the mechanism used to generate the photon mass is a Palatini-like formulation in the presence of the CFJ term. Interestingly, as we show in section III, we also find a photon mass proportional to the magnitude of the background vector.

Models with the presence of Lorentz-violating mass terms have been investigated before and present remarkable peculiarities. Some of these particularities were pointed out in \cite{lv-mass1} and \cite{lv-mass2}, in which a mass term $-(1/2)m^2 A_j A^j$ in electrodynamics was considered, being $j$ a spatial index. The gauge field, in this case, has two massive degrees of freedom, but the static force between charged particles is Coulomb-like. In \cite{lv-mass3}, it was investigated the possibility of radiatively generating a Lorentz-breaking mass for the photon in second order in the Lorentz-breaking vector. It was also carried out an analysis of more general mass terms, and the possibility of existence of superluminal modes in such cases was showed. A Stueckelberg lagrangian for massive photons in a generalized $R_\xi$ gauge was studied in \cite{lv-mass4}. Lorentz-breaking mass terms generated by spontaneous gauge symmetry breaking in a Lorentz-violating gauge-Higgs model  were investigated in \cite{lv-mass5}. In \cite{lv-mass6}, some aspects of this kind of gauge-symmetry breaking were focused in a study of dual models.

In this section, we obtained our model with an unusual Lorentz-breaking mass term with the use of a Palatini-like formulation of a CPT-odd QED containing a Chern-Simons-like part. This procedure is a firs-order formalism. We then discussed some general aspects of the obtained model. The treatments we are going to perform from now on are the ones usually adopted in field theory, which characterize a second-order formalism. In this sense, in this paper we have a mixture of first- and second-order formalisms, which is a new feature in the study of the CPT-odd Electrodynamics. Besides, we stress that the model to be studied in the next sections, which have the gauge sector of eq. (\ref{mass-lagrang}), is not the well-known one of eq. (\ref{CPTodd}). This new form of CPT-odd Electrodynamics motivates the studies which are carried out in the next two sections: we go deeper in the analysis of classical (section \ref{sec-propagator}) and quantum (section \ref{sec-quantum}) properties of the model.

\section{Classical propagator analysis}
\label{sec-propagator}
It is important to note that the sign of $\kappa^{2}$ might have strong implications in the gauge model of (\ref{mass-lagrang}). In particular, it is well known that consistency conditions concerning the issues of unitarity and causality are  directly related to the spacetime nature of the vector $\kappa^{\mu}$, that is, if it is spacelike, lightlike or timelike \cite{Andrianov:1998wj,Adam:2001ma,Belich}. In this section, we address these questions in the context of the model with the photon-sector described by equation (\ref{mass-lagrang}). Let us start by studying the physical consistency of the gauge sector given by the Lagrangian density (\ref{mass-lagrang}). Throughout this paper we will use the flat spacetime signature $\eta_{\mu\nu} = (1,-1,-1,-1)$ of the Minkowski space. Considering this Lagrangian density in the action, if we use partial integration, we obtain
\be
{\cal L}_G=\frac 12 A^\mu {\cal O}_{\mu \nu} A^\nu,
\label{parts-integration}
\ee
with
\be
{\cal O}_{\mu \nu}=(\Box+\kappa^2)\theta_{\mu \nu} + \kappa^2 \omega_{\mu \nu} + 2S_{\mu \nu} - \Lambda_{\mu \nu}.
\ee
where $\theta _{\mu \nu }=\eta_{\mu \nu} - \frac{\partial_\mu \partial_\nu}{\Box}$ and $\omega _{\mu \nu }=\frac{\partial_\mu \partial_\nu}{\Box}$ are respectively the transverse and longitudinal spin-projector operators. Besides, we use the $\kappa$-dependent operators $S_{\mu \nu }=\varepsilon_{\mu \nu \sigma \lambda }\kappa^{\sigma }\partial^{\lambda }$ and $\Lambda_{\mu \nu}=\kappa_\mu \kappa_\nu$. In order to obtain a closed algebra, one needs to include a new operator given by $\Sigma_{\mu \nu}=\kappa_\mu \partial_\nu$. We then proceed to the calculation of the classical propagator of the gauge field, given by
\be
\Delta_{\mu \nu}=i\left({\cal O}^{-1}\right)_{\mu\nu},
\ee
following the same procedure as in reference \cite{Belich}.

After a lengthy but straightforward calculation we obtain, in momentum-space,
\bq
&&\Delta_{\mu \nu}=\frac{i}{D_1}\left\{(-k^2 +\kappa^2)\theta_{\mu \nu}+\frac{1}{\kappa^2 D_2}\left[D_1 D_2
-\lambda^2\left(k^2(k^2-\kappa^2)-4D_2\right)\right]\omega_{\mu \nu} \nonumber \right. \\
&& \left. -2 i S_{\mu \nu} + \frac{1}{D_2}\left[\kappa^2(-k^2 + \kappa^2)-4D_2\right]\Lambda_{\mu \nu}
-\frac{\lambda(-k^2+\kappa^2)}{D_2}\left(\Sigma_{\mu \nu}+\Sigma_{\nu \mu}\right)\right\},
\label{propagator}
\eq
in which
\be
D_1=(k^2+\kappa^2+2\lambda)(k^2+\kappa^2-2\lambda)
\ee
and
\be
D_2=\lambda^2-\kappa^2k^2,
\label{completo-propagator}
\ee
with $\lambda=\kappa^\mu k_\mu$.

Once the gauge-field propagator is known, we are ready to discuss the particle content of the model. The elementary stable particles displayed in the spectrum of a model should appear as the poles of the field propagator. However, there are issues like causality and unitarity that have to be analyzed once the poles have been identified. This matter will be next discussed. With this purpose, we split our discussion into two cases: spacelike and timelike $\kappa^\mu$. In the analysis that follows, we choose particular frameworks which turns the calculation simpler without loss of generality when the question is the physical consistency of the model. However, some conditions we obtain are classical solutions in specific frameworks, which does not spoil the quantization of the model. As we know from basic Quantum Mechanics, the quantization, for example, by the path integral formalism takes into account all the possible trajectories.

\subsection{The case of $\kappa^\mu$ spacelike}

We begin by looking at the dispersion relations. For the denominator $D_1$, we have
\be
k^2+\kappa^2\pm 2\lambda=0,
\ee
which, if we have a purely spacelike background vector $\kappa^\mu=(0,\vec{\kappa})$, give us
\be
\omega^2=\left|\bold{k} \pm \vec{\kappa}\right|^2.
\ee
Since the model is massive, we can always go to the particle rest frame, with $\bold{k}=\vec{0}$. We then get, for the two dispersion relations,
\be
\omega^2=|\vec{\kappa}|^2.
\ee
It represents a particle with a positive definite mass which, in principle, propagates with two degrees of freedom. The same analysis in the dispersion relation coming from the denominator $D_2$ gives us $\omega^2=(1-\cos^2 \theta)|\bold{k}|^2$, being $\theta$ the angle between the momentum and the external vector $\vec{\kappa}$. It corresponds to a massless mode.

We now study the propagating wave. We set, without loss of generality, our external spacelike vector as given by $\kappa^\mu=(0,0,0,t)$. In this case, we have $\kappa^2=-t^2$ and $\lambda=-tk_z$. The propagator presents three poles. For the general denominator $D_1$, we have the poles $k_0^2\equiv m_1^2= k_x^2+k_y^2+(k_z+t)^2$ and $k_0^2\equiv m_2^2= k_x^2+k_y^2+(k_z-t)^2$. On the other hand, the denominator $D_2$, which is present in some of the terms, provides the pole $k_0^2\equiv\tilde{m}^2= k_x^2+k_y^2$.
Before analyzing the residue in the poles, we would like to check under which conditions the gauge-field propagates for each one of the modes. Let us write the field equation in momentum space:
\be
(-k^2+\kappa^2)A_\nu+k_\mu A^\mu k_\nu +2i \varepsilon_{\mu \nu \alpha \beta}\kappa^\alpha k^\beta A^\mu -\kappa_\mu A^\mu \kappa_\nu=0.
\ee
If we contract $k^\nu$ with the above equation, we obtain the gauge condition,
\be
k_\nu A^\nu= \left( \frac{\kappa_\mu k^\mu}{\kappa^2}\right) \kappa_\nu A^\nu,
\label{gauge}
\ee
such that the field equation yields
\be
(-k^2+\kappa^2)A_\nu+ \left[\left( \frac{\kappa_\mu k^\mu}{\kappa^2}\right)k_\nu-\kappa_\nu\right]\kappa_\mu A^\mu +2i \varepsilon_{\mu \nu \alpha \beta}\kappa^\alpha k^\beta A^\mu=0,
\ee
which for our spacelike $\kappa^\mu$ is written as
\be
-(k^2+t^2)A_\nu- \left(k_z k_\nu +t^2 \delta_\nu^3 \right)A_z +2it \varepsilon_{\mu \nu 3 \beta}k^\beta A^\mu=0.
\ee
We choose our coordinate system such that $A^\mu=(\phi,A_x,0,A_z)$, providing that $A_x$ is the $\bold{A}$ component orthogonal to $\vec{\kappa}$. Taking the third component of the field equation, we obtain
\be
(k^2+k_z^2)A_z=0.
\ee
Considering the $z$-component of the $A^\mu$ field is non null, we have $k_0^2=k_x^2+k_y^2$, which imposes some restrictions on the components of the momentum.

When the pole $k_0^2=\tilde m^2$ is considered along with the field equations, we obtain the constraints $k_z=0$ and $\phi=A_x=0$. This means that the pole $k_0^2=\tilde m^2$ is to be associated with a photon polarized parallel to $\vec{\kappa}$ and  propagating in a direction orthogonal to this Lorentz-breaking vector. Besides, it is a transversal mode of propagation.

For the poles $k_0^2=m_1^2$ and $k_0^2=m_2^2$, the relations $k_x^2+k_y^2=k_x^2+k_y^2+(t \pm k_z)^2$ leave us with the conditions $k_z=-t$ and $k_z=t$ for the modes corresponding to the poles $k_0^2=m_1^2$ and $k_0^2=m_2^2$, respectively. It is interesting to note that while the massless pole $k_0^2=\tilde m^2$ is associated to a transversal mode of propagation with polarization parallel to the external vector, the modes $k_0^2=m_1^2$ and $k_0^2=m_2^2$ are such that the $z$-component of the momentum compensates the magnitude of the background vector. However, it is important to enforce that the spatial scalar product $\vec \kappa \cdot \bold k$ is not invariant under Lorentz transformation. The above result is a feature of a framework for which $\kappa^\mu$ is purely spacelike.

To infer about the physical nature of the poles, we have to investigate issues like unitarity and causality. Unitarity at tree-level can be investigated analyzing the propagator saturated by conserved currents,
\be
{\cal SP}=J^\mu\Delta_{\mu \nu}J^\nu.
\ee
The current conservation in momentum-space is written as $k_\mu J^\mu=0$. Unitarity requires that the residue of the saturated propagator in a physical pole is positive definite (see \cite{Cas-Man} and \cite{Veltman}). This requirement can be checked by calculating the residue matrix in the pole for the complete propagator and, then, verifying if its eigenvalues are positive definite. Here we opt for analyzing directly the saturated propagator, which is given by
\be
{\cal SP}=\frac{i}{D_1}\left\{(-k^2+\kappa^2)J^2+\frac{1}{D_2}\left[\kappa^2(-k^2+\kappa^2)-4D_2\right](J\cdot b)^2\right\}.
\ee
For our spacelike $\kappa^\mu$, we get
\be
\label{SP-spacelike}
{\cal SP}=-\frac{i}{(k_0^2-m_1^2)(k_0^2-m_2^2)}\left\{(k^2+t^2)J^2+\frac{t^2J_z^2}{k_0^2-\tilde m^2}\left[-(k^2+t^2)+4(k_0^2-\tilde m^2)\right]\right\}.
\ee

We carry out below the analysis of the particular cases we have just obtained.

\subsubsection{$k_0^2=\tilde m^2$, with $k_z=0$}

When $k_z=0$, we get $m_1^2=m_2^2=\tilde m^2+t^2$ and $k^2+t^2=k_0^2-\tilde m^2+t^2$. The saturated propagator, in this case, reads
\be
{\cal SP}=-\frac{i}{(k_0^2-\tilde m^2-t^2)^2}\left\{(k_0^2-\tilde m^2-t^2)J^2
+\frac{t^2 J_z^2}{k_0^2-\tilde m^2}\left[(k_0^2-\tilde m^2-t^2)+4(k_0^2-\tilde m^2)\right]\right\}.
\ee

The calculation of the residue in the pole $k_0^2=\tilde m^2$ is straightforward and yields
\be
{\cal R}_{k_0^2=\tilde m^2}({\cal SP})=J_z^2,
\ee
which is positive definite.

\subsubsection{$k_0^2=m_1^2$, with $k_z=-t$}

When we look at eq. (\ref{SP-spacelike}), we see that, when $k_z=-t$, and consequently $m_1^2=\tilde m^2$, the saturated propagator apparently exhibits a dangerous double pole. However, it occurs an interesting cancelation of one power of $k_0^2- \tilde m^2$, since, in this case, we have $k_0^2-|\bold{k}|^2 +t^2=k_0^2- \tilde m^2$. The saturated propagator, in this particular situation, takes the form
\be
{\cal SP}=-\frac{i}{(k_0^2-\tilde m^2)(k_0^2-\tilde m^2-4t^2)}\left\{(k_0^2-\tilde m^2)J^2+3t^2 J_z^2\right\}.
\ee

The calculation of the residue in the pole $k_0^2=m_1^2=\tilde m^2$ yields
\be
{\cal R}_{k_0^2=m_1^2}({\cal SP})=\frac 34 J_z^2,
\ee
which is positive definite.

\subsubsection{$k_0^2=m_2^2$, with $k_z=t$}

We now consider the pole $k_0^2=m_2^2$ in the particular situation in which $k_z=t$. Again the pole collapses to $\tilde m^2$, and the possibility of a double pole arises. However, as in the previous case, a factor of $k_0^2-\tilde m^2$ is canceled and we are left with a simple pole. We proceed to identical calculations as in the last subsection to obtain
\be
{\cal R}_{k_0^2=m_2^2}({\cal SP})= \frac 34 J_z^2,
\ee
which is positive definite.

We observe that, in the case of a spacelike background vector $\kappa^\mu$, we have three simple poles, $k_0^2=m_1^2$, $k_0^2=m_2^2$ and $k_0^2=\tilde m^2$, which contribute each one with one degree of freedom of the propagating gauge field. Interestingly, the propagating modes are constrained to collapse to the pole $k_0^2=\tilde m^2$.

Concerning microcausality, models with Lorentz-breaking mass terms are known to, in some situations, accommodate supraluminal modes of propagation, as well discussed in \cite{lv-mass3}. However, this is not the case here, since, as we obtained, the component of the momentum of the propagating wave which is parallel to $\vec{\kappa}$ always compensates the breaking vector.

\subsection{The case of $\kappa^\mu$ timelike}

We set our external timelike vector as given by $\kappa^\mu=(t,0,0,0)$, so that we have $\kappa^2=t^2$ and $\lambda=tk_0$. The denominator $D_1$ becomes
\be
D_1=(k_0^2-m_1'^2)(k_0^2-m_2'^2),
\ee
with $m_1'^2=(|\bold{k}|+t)^2$ and $m_2'^2=(|\bold{k}|-t)^2$, while $D_2=t^2|\bold{k}|^2$. In this case, $D_2$ provides a massive pole, since $|\bold{k}|=0$ for $D_2=0$. This atypical situation is caused by the choice of a framework such that $\kappa^\mu=(t,0,0,0)$, which, by coincidence, is the particle rest frame.

As before, we begin our analysis by the momentum-space field equation which, for our timelike $\kappa^\mu$, reads
\be
(-k^2+t^2)A_\nu+(k_0 k_\nu -t^2 \delta^0_\nu)\phi + 2it \varepsilon_{\mu \nu 0 \beta}k^\beta A^\mu=0.
\ee
For $\nu=0$, we get
\be
|\bold{k}|^2 \phi=0,
\ee
which sets $|\bold{k}|=0$ or $\phi=0$. On the other hand, the spatial indices give us the vectorial equation
\be
(k^2-t^2)\bold{A}+2it(\bold{k}\times\bold{A})=0.
\label{vector}
\ee

A possible solution would be a momentum parallel to $\bold{A}$, with $k^2-t^2=0$. However, if $m_1'^2$ or $m_2'^2$ is substituted in the above equation, it is obtained $|\bold{k}|=0$. The same for the pole coming from $D_2$. So our framework choice is such that the particle is at rest. Moreover, a timelike $\kappa^\mu$ is problematic, since the double pole which arises from the conditions above does not cancel as in the case the background vector is spacelike.

A comment is in order. The unitarity considered here is not in the sense of a self-adjoint extension, but rather in the framework of the Hilbert space of particle states.

\section{One loop radiative corrections}

After discussing classical aspects of the propagating modes of the vectorial field $A^\mu$, we turn our attention to the quantum corrections to our model. Since the mass term which arises due to the Palatini formulation with the contribution of the CFJ term is second order in $\kappa^\mu$, it is recommendable to investigate if, up to second order in this vector, the renormalization of the model is not affected.  The complete action for our Lorentz- and CPT-violating model is given by
\begin{equation}
S = \int d^4x\left\{\mathcal{L}_G +\bar \psi \left(i \Dslash-m-e \Aslash - \bslash\gamma_{5}\right)\psi\right\}.
\label{Model}
\end{equation}

Before starting our calculations, we go through some background literature. We have the fermion coupled to an axial vector $b^\mu$. It is known that an axial vector coupled to a massive fermion describes an antisymmetric torsion. Consistency conditions for preserving unitarity and renormalizability were found in \cite{torsion1}, which impose severe restrictions in this kind of model. Besides, many quantum aspects of torsion theory were reviewed in \cite{torsion2}. The present model, however, deals with a constant background which has been shown to preserve all the desirable features of a quantum theory. Concerning our mass term of the type $\frac 12 A^\mu M_{\mu \nu} A^\nu$, several recent works in the context of curved spacetime have dealt with the quantization of Abelian models possessing a similar term, which is called nonminimal \cite{Curv-space1}-\cite{Curv-space4}. In curved spacetime, this kind of term provides a nonminimal coupling between vector and gravitational fields. This fact causes difficulties with renormalizability, for example. Some procedures are applied to take care of these problems, like the formalism of Stueckelberg. In the present work, we treat an Abelian model in flat spacetime, which accommodates Lorentz-violating contributions. Our gauge-breaking term is completely built from the Lorentz breaking parameter $\kappa^\mu$, which brings new features. We emphasize that our aim is not to prove renormalizability of the model, which certainly needs some care that is left for a future study, but to investigate its one-loop consistency up to second order in the Lorentz breaking parameters.

\begin{figure}[h]
\begin{center}
\includegraphics[scale=0.5]{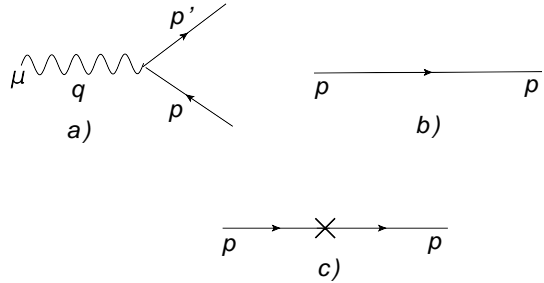}
\caption{(a)  the trilinear vertex of QED; (b) the usual fermion propagator; and (c) the Lorentz-breaking insertion in the fermion line.}
\label{feynman-rules}
\end{center}
\end{figure}

First, we note that there is no modification here in the interaction between the fields $A^\mu$ and $\psi$ of QED, which is described by the conventional minimal coupling $-ie\gamma_\mu$ (Fig. \ref{feynman-rules}a). Considering the Lorentz-violating terms in (\ref{Model}), since the $b^\mu$ vector appears in only one term of the model, we treat it as a perturbation of the QED Lagrangian, so that it is used as a vertex insertion within the Feynman diagrams of QED (Fig. \ref{feynman-rules}c):
\begin{equation}
V_b=-ib_\mu \gamma^\mu \gamma_{5}.
\label{odd-fermion-vertex}
\end{equation}

\begin{figure}[h]
\begin{center}
\includegraphics[scale=0.5]{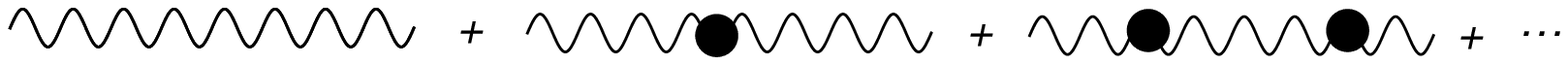}
\caption{Diagrammatic representation of the expansion of the classical gauge propagator (\ref{propagator}) up to second order in $\kappa^{\mu}$. The first diagram represents the usual photon propagator, whereas the second and third terms correspond, respectively, to $\Delta_{\mu \nu}^{(1)}$ and $\Delta_{\mu \nu}^{(2)}$.}
\label{sum-insertions}
\end{center}
\end{figure}

Concerning the terms which contain the Lorentz-breaking vector $\kappa^\mu$, they might also be treated as perturbations of the original Lagrangian density. However, since we would have to consider two different kinds of insertions (one at first order and another at second order in $\kappa^\mu$) for the same background vector, we opt to use the propagator for the gauge field up to second order in $\kappa^\mu$. With this aim, we expand the complete photon propagator of equation (\ref{propagator}) in the Lorentz-violating parameter $\kappa^\mu$,
\bq
\Delta^{\mu\nu}(p)&=& \Delta^{\mu\nu}(p)\mid_{\kappa=0}+\frac{\partial\Delta^{\mu\nu}(p)}{\partial\kappa^{\rho}}\mid_{\kappa=0}\kappa^{\rho}
+\frac{1}{2}\frac{\partial^{2}\Delta^{\mu\nu}(p)}{\partial\kappa^{\rho}\partial\kappa^{\sigma}}\mid_{\kappa=0}\kappa^{\rho}\kappa^{\sigma}
+\mathcal{O}(\kappa^{3})
\eq
to obtain
\bq
&&\Delta^{\mu \nu}=\Delta^{\mu\nu}_{{\tiny QED}}(p)-\frac{2}{p^{4}}\epsilon^{\mu\nu\alpha\lambda}\kappa_{\alpha}p_{\lambda}
 +\frac{i}{p^{6}}\left\{2p^2\left(\kappa^{2}\eta^{\mu\nu}-\kappa^{\mu}\kappa^{\nu}\right)\right. \nonumber \\
&&\left. - 4\eta^{\mu\nu}\left(\kappa\cdot p\right)^{2}-4p^{\mu}p^{\nu}\kappa^{2}+4(\kappa\cdot p)\left(p^{\mu}\kappa^{\nu}+p^{\nu}\kappa^{\mu}\right)\right\} + \mathcal{O}(\kappa^{3})  \nonumber\\
&& \equiv \Delta^{\mu \nu(0)}+ \Delta^{\mu \nu(1)}+\Delta^{\mu \nu(2)}+ \mathcal{O}(\kappa^{3}),
\label{expansion}
\eq
in which $\Delta_{\mu \nu}^{(0)}$, the zeroth-order term in $\kappa^\mu$, is the photon propagator of the traditional QED, $\Delta_{\mu \nu}^{(1)}$  is the contribution to the propagator in first order in $\kappa^\mu$ and so on. The expansion can be diagrammatically represented by Fig. \ref{sum-insertions}.

\label{sec-quantum}

Finally, since we are treating the axial term in the fermionic sector as a perturbation, we are left with the usual fermion propagator (Fig. \ref{feynman-rules}b), given by
\begin{equation}
S^{QED}(p) = \frac{i}{\pslash - m}.
\label{propag2}
\end{equation}

\label{sec-quantum}

Here we are interested in the one-loop Lorentz-violating contributions to the two-point functions of the photon and fermion fields, since the three-point function will not present new divergent integrals due to the Lorentz-breaking. Figures \ref{two-point-gauge} and \ref{electron-3vertex} show the Feynman diagrams up to second order in the Lorentz-violating vectors $\kappa^\mu$  and $b^\mu$.

\subsection{Two point function of the $A^\mu$ field}

\begin{figure}[h]
\begin{center}
\includegraphics[scale=0.5]{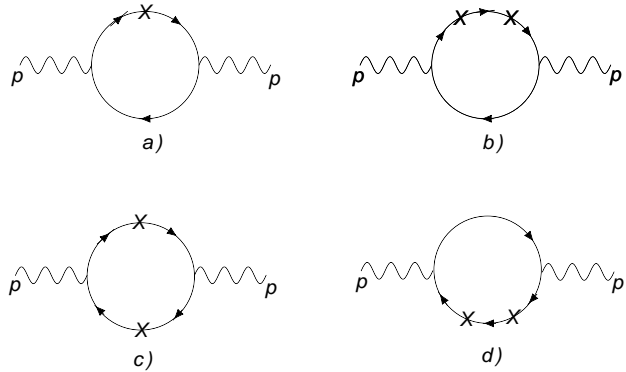}
\caption{Feynman diagrams for the photon two-point function with insertions of the axial vertex (\ref{odd-fermion-vertex}), up to the second order.}
\label{two-point-gauge}
\end{center}
\end{figure}

Let us start with the diagrams in Fig. \ref{two-point-gauge}, which represent Lorentz-violating contributions to the vacuum polarization tensor at one-loop order. For the photon two-point function, the first order contribution in $b^\mu$, represented by the diagram in Fig. \ref{two-point-gauge}a, is the well-known CFJ term, whose quantum generation has been extensively discussed in the last two decades. We focus on the second order contribution.

Thus, we next consider the  diagrams in Figures \ref{two-point-gauge}b, \ref{two-point-gauge}c and \ref{two-point-gauge}d, which have two insertions of the vertex  (\ref{odd-fermion-vertex}). The expressions corresponding to these diagrams can be derived by using the Feynman rules presented in the last section. We have, respectively
\begin{eqnarray}
\Pi^{(3b)}_{\mu\nu}&=&-e^{2}\int \frac{d^{4}k}{(2\pi)^{4}}\frac{{\rm Tr}\Big\{\gamma^{\nu}[(\pslash-\kslash) + m)]\gamma^{\mu}(\kslash + m)\bslash\gamma_{5}(\kslash+m)\bslash\gamma_{5}(\kslash + m)\Big\}}{(k^{2}-m^{2})^{3}[(k-p)^{2}-m^{2}]},
\label{diagram2a}
\end{eqnarray}
\begin{eqnarray}
\Pi^{(3c)}_{\mu\nu}&=&-e^{2}\int \frac{d^{4}k}{(2\pi)^{4}}\frac{{\rm Tr}\Big\{\gamma^{\nu}[(\pslash-\kslash) + m)]\bslash\gamma_{5}[(\pslash-\kslash) + m]\gamma^{\mu}(\kslash + m)\bslash\gamma_{5}(\kslash+m)\Big\}}{(k^{2}-m^{2})^{2}[(k-p)^{2}-m^{2}]^{2}}
\label{diagram2b}
\end{eqnarray}
and
\begin{eqnarray}
\Pi^{(3d)}_{\mu\nu}&=&-e^{2}\int \frac{d^{4}k}{(2\pi)^{4}}\frac{{\rm Tr}\Big\{\gamma^{\nu}[(\pslash-\kslash) + m)]
\bslash \gamma_5 [(\pslash-\kslash) + m)]\bslash \gamma_5 [(\pslash-\kslash) + m)]\gamma^{\mu}(\kslash + m)\Big\}}{(k^{2}-m^{2})[(k-p)^{2}-m^{2}]^3}.
\label{diagram2d}
\end{eqnarray}

After computing the three contributions, we obtain the following finite result
\bq
&&\Pi^{bb}_{\mu \nu}=-i\frac{e^2}{2 \pi^2}\left\{\frac 16 b^2\eta_{\mu \nu}+ b^2 \eta_{\mu \nu}\int_0^1dx(2-3x)x\ln\left(\frac{H^2}{m^2}\right) \right. \nonumber \\
&&\left.-m^2b^2 \eta_{\mu \nu} \int_0^1dx\frac{x(1-x)}{H^2}+4m^2b^2(p_\mu p_\nu-p^2 \eta_{\mu \nu})\int_0^1dx\frac{x(1-x)^2}{(H^2)^2}
\right. \nonumber \\
&& \left. +4m^2p^2b_\mu b_\nu \int_0^1dx\frac{x^2(1-x)}{(H^2)^2}-2m^2(p\cdot b)(p_\mu b_\nu+ p_\nu b_\mu)\int_0^1dx\frac{x(1-x)}{(H^2)^2}
\right. \nonumber \\
&&\left. +2m^2 (p\cdot b)^2 \eta_{\mu \nu} \int_0^1dx\frac{x^2(1-x^2)}{(H^2)^2}
-4(p\cdot b)^2\eta_{\mu \nu}\int_0^1dx\frac{x^2(1-x)(1-2x)}{H^2}\right. \nonumber \\
&&\left.+b^2p^2 \eta_{\mu \nu}\int_0^1dx\frac{x^3(1-x)}{H^2}+2p^2(p\cdot b)^2 \eta_{\mu \nu}\int_0^1dx\frac{x^4(1-x)^2}{(H^2)^2}\right\},
\eq
in which $H^2=-p^2x(1-x)+m^2$. This result can be organized in a manifest gauge-invariant form. For simplicity, we write the result in powers of $p^2$ as
\be
\Pi_{\mu \nu}=-i\frac{e^2}{6\pi^2m^2}\left[1+\frac 25 \frac{p^2}{m^2} + {\cal O}\left(\frac{p^4}{m^4}\right)\right]T_{\mu \nu},
\ee
with
\be
T_{\mu \nu}=b^2(p_\mu p_\nu-p^2 \eta_{\mu \nu})+p^2b_\mu b_\nu +(p \cdot b)^2\eta_{\mu \nu}-(p \cdot b)(p_\mu b_\nu+p_\nu b_\mu).
\ee

We see from the tensor $T_{\mu \nu}$ that the second-order correction in $b^\mu$ has a Maxwell part and a contribution with the aether form $(b_\mu \bar F^{\mu \nu})^2$, with $\bar F_{\mu \nu}=\partial_\mu A_\nu-\partial_\nu A_\mu$. Besides, we got higher-order derivatives of these terms. The terms in $p^2$ are in agreement with the ones obtained in \cite{Bonneau} and \cite{albert1}. It is interesting to note that, at one-loop order, there is no correction to the Lorentz-violating mass term generated at the classical level for the vectorial field $A^\mu$. This occurs due to a cancelation of divergencies coming from different contributions and the use of a regularization prescription which sets surface terms to zero. Here we have used Implicit Regularization \cite{implicit}, but it also happens for any gauge-invariant technique. If the use of a gauge-invariant regularization method is relaxed, a finite correction to the Lorentz-breaking mass term would be obtained (see \cite{Altschul} and \cite{EPJC-Scarp}).

\subsection{Two point function of the $\psi$ field}

 \begin{figure}[h]
\begin{center}
\includegraphics[scale=0.5]{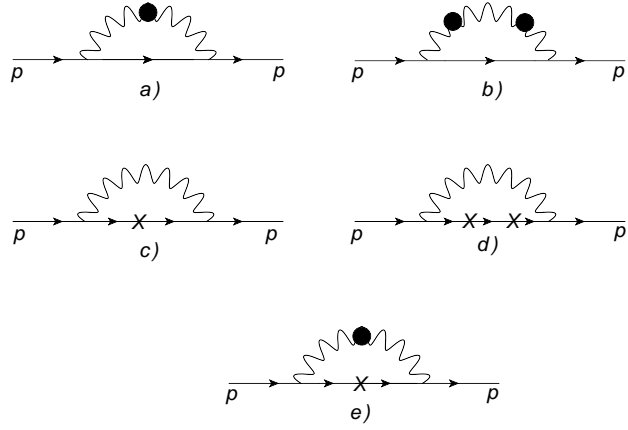}
\caption{Feynman diagrams for the fermion two-point function with one and two insertions of the Lorentz-breaking vectors.}
\label{electron-3vertex}
\end{center}
\end{figure}

Let us now compute the Lorentz-breaking corrections to the two-point function for the fermionic field. The Feynman graphs which represent the contributions up to second-order in the background vectors are presented in Fig. \ref{electron-3vertex}.

It is worth to note that only the first-order corrections (Figures \ref{electron-3vertex}a and \ref{electron-3vertex}c) will deliver divergent integrals. The corresponding amplitudes are given below

\begin{eqnarray}
\Sigma_{4a}&=&2ie^{2}\kappa_{\sigma}\epsilon^{\sigma\mu\rho\nu}\int \frac{d^{4}k}{(2\pi)^{4}}\frac{\gamma_{\nu}[(\pslash-\kslash)+m]\gamma_{\mu}k_{\rho}}{k^{4}[(k-p)^{2}-m^{2}]}
\label{3a}
\end{eqnarray}
and
\begin{eqnarray}
\Sigma_{4c}&=&-e^{2}\int \frac{d^{4}k}{(2\pi)^{4}}\frac{\gamma_{\nu}(\kslash + m)\bslash\gamma_{5}(\kslash + m)\gamma^{\nu}}{(k^{2}-m^{2})^{2}(k-p)^{2}}.
\label{3c}
\end{eqnarray}
It is important to remember that in the graph of Fig. \ref{electron-3vertex}a, the photon line with one insertion means the term $\Delta_{\mu \nu}^{(1)}$ in the expansion of the photon propagator. These two contributions furnish the results
\bq
&&\Sigma_{4a}=e^2\left\{-3 \kappaslash I_{log}(\Lambda^2)
+3 \frac{i}{16 \pi^2} \kappaslash \int_0^1 dx \ln \left(\frac{M^2}{\Lambda^2}\right) \right.\nonumber \\
&&\left.-\frac{i}{4\pi^2}(\kappaslash \pslash -\kappa \cdot p)\left[m\int_0^1 dx\frac{x(1-x)}{M^2}+\pslash\int_0^1\frac{(1-x)^2x}{M^2}\right]\right\}\gamma_5
\eq
and
\bq
&&\Sigma_{4c}=e^2\left\{-\bslash I_{log}(\Lambda^2)+\frac{i}{16 \pi^2}\bslash \int_0^1 dx \ln \left(\frac{M^2}{\Lambda^2}\right) \right. \nonumber \\
&&\left. -\frac{i}{4 \pi^2}(p \cdot b)\pslash \int_0^1 dx \frac{(1-x)^2x}{M^2}+\frac{i}{4 \pi^2}m^2 \bslash \int_0^1 dx \frac{x}{M^2}\right\}\gamma_5,
\eq
where
\be
I_{log}(\Lambda^2)=\int \frac{d^4k}{(2\pi)^4}\frac{1}{(k^2-\Lambda^2)^2},
\ee
\be
M^2=x(m^2-p^2(1-x))
\ee
and $\Lambda^2$ is a mass scale introduced in the process of Implicit Regularization \cite{implicit,implicit1,implicit2}. We perform an expansion in powers of the external momentum $p^\mu$ and obtain, up to the first-order,
\bq
&&\Sigma_{4a}+\Sigma_{4c}=e^2\left\{-(3\kappaslash+\bslash) I_{log}(\Lambda^2)
+ \frac{i}{16 \pi^2}\left\{(3\kappaslash+\bslash)\ln\left(\frac{m^2}{\Lambda^2}\right) +3(\bslash-\kappaslash)\right. \right. \nonumber \\
&&\left. \left.-2\frac{(\kappaslash \pslash -\kappa \cdot p)}{m}\right\}\right\}\gamma_5 +{\cal O}(p^2).
\eq

The contributions of second-order in the Lorentz-breaking parameters are represented by the graphs of Figures \ref{electron-3vertex}b, \ref{electron-3vertex}d and \ref{electron-3vertex}e and are given by
\be
\Sigma_{4b}=e^{2}\int \frac{d^{4}k}{(2\pi)^{4}}\frac{\gamma^\nu(\pslash-\kslash+m)\gamma^\mu}{[(k-p)^{2}-m^{2}]}\frac{\Delta^{(2)}_{\mu \nu}}{i},
\ee
\be
\Sigma_{4d}=-e^2 \int \frac{d^{4}k}{(2\pi)^{4}}\frac{\gamma^\mu(\kslash+m)\bslash \gamma_5 (\kslash+m)\bslash \gamma_5 (\kslash+m)\gamma_\mu}
{(k^2-m^2)^3(p-k)^2}
\ee
and
\be
\Sigma_{4e}=2ie^2\epsilon^{\mu \nu \alpha \beta}\kappa_\alpha\int \frac{d^{4}k}{(2\pi)^{4}}\frac{\gamma_\mu(\kslash+m)\bslash \gamma_5 (\kslash+m)\gamma_\nu (p-k)_\beta}{(k^2-m^2)^2(p-k)^4}.
\ee
All these Feynman integrals will deliver finite results, which are given by
\be
\Sigma_{4b}=\frac{i}{4 \pi^2}e^{2}\left\{\left[2\kappaslash(p \cdot \kappa)-\pslash\kappa^{2}\right]\int_{0}^{1}dx\frac{(1-x)x}{M^2}- \pslash(p \cdot \kappa)^{2}\int_{0}^{1}dx\frac{(1-x)^{2}x^{3}}{M^4}\right\},
\ee
\bq
&&\Sigma_{4d}=\frac{i}{8 \pi^2}e^2\left\{2mb^2\left[2m^2\int_0^1 dx \frac{x^2}{M^4}-\int_0^1 dx\frac{x(3-x)}{M^2}\right]
+2(p \cdot b)\bslash \int_0^1 dx \frac{x(1-x)^2}{M^2} \right.\nonumber \\
&&\left. +b^2\pslash\left[\int_0^1 dx \frac{x(1-x)^2}{M^2}-2m^2\int_0^1 dx \frac{x^2(1-x)}{M^4}\right] \right. \nonumber \\
&&\left.+2(p \cdot b)^2\int_0^1 dx \frac{x^2(1-x)^2 \left[(1-x)\pslash-2m\right]}{M^4}\right\}
\eq
and
\bq
&&\Sigma_{4e}=-\frac{1}{8 \pi^2}e^2\epsilon^{\alpha \nu \beta \sigma} \kappa_\alpha \gamma_\sigma
\left\{-m(\gamma_\beta \bslash -b_\beta)\int_0^1 dx \frac{x(1-x)}{M^2} -2m^2 \bslash p_\beta \int_0^1 dx \frac{(1-x)x^2}{M^4} \right. \nonumber \\
&& \left. -2m[\pslash \bslash -(p \cdot b)]p_\beta \int_0^1 dx \frac{x^2(1-x)^2}{M^4}
+[(p \cdot b)\gamma_\beta+\pslash b_\beta]\int_0^1 dx \frac{x(1-x)^2}{M^2} \right. \nonumber \\
&&\left. +2(p \cdot b)\pslash p_\beta \int_0^1 dx \frac{x^2(1-x)^3}{M^4} +\bslash p_\beta \int_0^1 dx \frac{x(1-x)^2}{M ^2}\right\} \gamma_\nu \gamma_5.
\eq

We again carry out an expansion in powers of $p^\mu$, up to the first-order, to write
\bq
&&\Sigma_{4b}+\Sigma_{4d}+\Sigma_{4e}= \frac{i}{8 \pi^2} \frac{e^2}{m}\left\{ -b^2-3(b \cdot \kappa)+\frac{1}{3m}\left\{
\left[-3\kappa^2-2b^2+4(\kappa \cdot b)\right]\pslash \right. \right.\nonumber \\
&&\left. \left.+2 p \cdot (3\kappa +2 b)\kappaslash +2p \cdot(b-\kappa)\bslash \right\}\right\} + {\cal O}(p^2).
\eq

Let us now comment on the results of this section. Considering the first-order quantum corrections to the fermion two-point function, we found that both contributions, in $b^\mu$ and in $\kappa^\mu$, are divergent. Actually, a classical  Lorentz-violating extension of QED with only the CFJ-term (that is, with $b^{\mu}=0$) would give us an inconsistent  quantum field theory since a logarithmically divergent Lorentz-violating term in $\bar{\psi}\gamma_{5}\gamma^{\mu}\psi$ is obtained  by radiative correction from the two-point Green's function of the fermion field $\psi$. This fact was already known from \cite{Kost-ren}, in which the one-loop renormalization of the extended QED at first order in the Lorentz-breaking parameters was performed. The second-order corrections, on the other hand, are finite and contribute to the kinetic- and mass-terms of the $\psi$-field.

\section{Conclusion}
\label{sec-conclusion}

We presented a model which incorporates CPT-odd Lorentz-breaking terms in both gauge and fermionic sectors. In this model, the gauge sector is written in the Palatini-like form, such that $A^{\mu}$ and $F^{\mu\nu}$ are treated as independent fields. This formulation naturally relaxes gauge-invariance at the classical level in a first view and gives rise to a CPT-even Lorentz-violating mass term for the photon field. In other words, a Palatini-like formulation, with the contribution of Lorentz violation, induced a Lorent-breaking mass term which, in turn, causes the violation of gauge symmetry. It is relevant to enforce the role of the CFJ term is this process, since if we have $\kappa^\mu=(0,0,0,0)$ the gauge symmetry is restored. Therefore, we have an interesting case in which Lorentz-symmetry violation causes the breaking of gauge invariance. Although this breakdown of gauge invariance can cause undesirable problems to the model, we believe a procedure like the one developed in \cite{lv-mass4} for applying Stueckelberg formalism to Lorentz-breaking theories could also be adopted in the present model.
It is possible that this model is obtained by gauge fixing a Lorentz-violating gauge-invariant Stueckelberg Lagrangian.

This new form of the gauge sector in the classical action was investigated through the analysis of the poles of the photon propagator for the case the background vector $\kappa^\mu$ is spacelike and timelike. For a spacelike $\kappa^\mu$, we have three simple poles which contribute each one with one degree of freedom of the propagating gauge field. For each mode, the field equations provide propagating waves with a momentum which compensates the background vector, such that microcausality is preserved. Interestingly, these restrictions force the propagating modes to collapse to the pole $k_0^2=\tilde m^2$. On the other hand, there are no acceptable modes of propagation of the gauge field for timelike $\kappa^\mu$. Besides, the study of the classical gauge action carried out in section III showed, for spacelike $\kappa^\mu$, a mass for the photon proportional to the magnitude of the spatial component of the background vector, which is in accordance with the results of \cite{Bonetti:2016vrq} and \cite{bonettiA}.

We also performed the calculation of one-loop quantum corrections to the two-point functions for both gauge and fermion fields up to the second-order in the Lorentz-violating vectors. Concerning the corrections to the photon line, the second-order correction in $b^\mu$ has a Maxwell part and a contribution with the aether-form $(b_\mu \bar F^{\mu \nu})^2$. Besides, we got higher-order derivatives of these terms. There are no corrections to the mass term at one-loop order. The first-order quantum corrections to the fermion two-point function enforced that a classical  Lorentz-violating extension of QED with only the CFJ-term would give us an inconsistent  quantum field theory since a logarithmically divergent Lorentz-violating term in $\bar{\psi}\gamma_{5}\gamma^{\mu}\psi$ is obtained. The second-order corrections to the fermion line, on the other hand, are finite, and contribute to the kinetic- and mass-terms of the $\psi$-field.

The renormalizability of the model at one-loop order is not affected by the mass term, although a complete one-loop renormalization of the model is out of the scope of the present work. It would be interesting to investigate in an upcoming work how the Lorentz-breaking mass term classically induced here would affect the renormalization and the Ward identities of the theory in higher-loop calculations. In this investigation, it would be interesting to adopt a Stueckelberg procedure to restore gauge invariance of the model.

\vspace{1.0cm}

\noindent {\bf Acknowledgments}

This work was partially supported by Conselho Nacional de Desenvolvimento Cient\'{\i}fico e Tecnol\'{o}gico (CNPq). The authors acknowledge Prof. J. A. Helay\"el-Neto for elucidating discussions.


\end{document}